\documentclass[runningheads,a4paper]{llncs}

\usepackage{times}
\usepackage{amssymb}
\usepackage{graphicx}
\usepackage{url}
\usepackage{multirow}

\usepackage{algorithm}
\usepackage{algorithmic}

\usepackage{amssymb}

\newcommand{\keywords}[1]{\par\addvspace\baselineskip
\noindent\keywordname\enspace\ignorespaces#1}

\begin{document}

\title{CML-TTS: A Multilingual Dataset for Speech Synthesis in Low-Resource Languages}

\titlerunning{CML-TTS: A Multilingual Dataset for Speech Synthesis in Low-Resource Languages}

\author{Frederico S. Oliveira \and Edresson Casanova \and Arnaldo Cândido Júnior \and Anderson S. Soares \and Arlindo R. Galvão Filho}


\authorrunning{Oliveira et al.}

\institute{UFG, Goiás - GO - Brazil \\
}

\index{Oliveira, Frederico}
\index{Casanova, Edresson}
\index{Cândido Júnior, Arnaldo}
\index{Soares, Anderson}
\index{Galvão Filho, Arlindo}

\toctitle{} \tocauthor{}

\maketitle

%
%
%
%
\begin{abstract}
In this paper, we present CML-TTS, a recursive acronym for CML-Multi-Lingual-TTS, a new Text-to-Speech (TTS) dataset developed at the Center of Excellence in Artificial Intelligence (CEIA) of the Federal University of Goias (UFG). CML-TTS is based on Multilingual LibriSpeech (MLS) and adapted for training TTS models, consisting of audiobooks in seven languages: Dutch, French, German, Italian, Portuguese, Polish, and Spanish. Additionally, we provide the YourTTS model, a multi-lingual TTS model, trained using 3,176.13 hours from CML-TTS and also with 245.07 hours from LibriTTS, in English. Our purpose in creating this dataset is to open up new research possibilities in the TTS area for multi-lingual models. The dataset is publicly available under the CC-BY 4.0 license\footnote{\url{https://freds0.github.io/CML-TTS-Dataset}}.
\end{abstract}

\section{Introduction}

Text-To-Speech (TTS) systems have received a lot of attention in recent years due to the great advance provided by the use of Deep Learning, which allowed the popularization of virtual assistants, such as Apple Siri \cite{Gruber2009}, Amazon Alexa \cite{purington2017alexa} and Google Home \cite{dempsey2017teardown}. Traditional TTS systems, according to \cite{Tachibana2017}, were composed of several specific modules, which are difficult to develop, such as a text analyzer, a grapheme-to-phoneme converter, a duration estimator, and an acoustic model \cite{Ze2013statistical}. Deep learning \cite{deeplrgoodfellow} allows the integration of all these modules into a single model, producing spectrograms from texts, with good performance and quality \cite{Casanova2022tts}. As examples, check \cite{kyle2017char2wav,Tachibana2017,Wang2017tacotron,Ping2017deepvoice3,Shen2018tacotron2,kim2020glow,valle2020flowtron,kim2021conditional,huang2022prodiff}.

The difficulty in training models based on Deep Learning is that these models, such as \cite{kyle2017char2wav,Wang2017tacotron,Tachibana2017,Ping2017deepvoice3,Shen2018tacotron2}, require a greater amount of data for training. For this reason, most current TTS models are designed for the English language \cite{Ping2017deepvoice3,Shen2018tacotron2,valle2020flowtron,kim2020glow,kim2021conditional,huang2022prodiff}, which is a language with many open resources. The main datasets available for training TTS models in the English language are Voice Cloning Toolkit (VCTK) \cite{Yamagishi2016Vctk}, LJSpeech \cite{Ito2017Ljspeech} and LibriTTS corpus \cite{Zen2019libritts}. VCTK \cite{Yamagishi2016Vctk} is a dataset comprising a total of 44 hours of recordings, with 109 native English speakers, in which each speaker reads approximately 400 sentences. In VCTK, recordings were made in a studio, with high quality, and a sampling rate equal to 48kHz. LJSpeech is a single-speaker reading dataset in English, which contains 24 hours of audiobook reading with a sampling rate equal to 22kHz, whose recordings come from the LibriVox project. LibriTTS is a dataset adapted from the LibriSpeech \cite{Panayotov2015LibrispeechAA} corpus, which contains 585 hours of speech at 24kHz, and is composed of 2,456 speakers. 

With the increasing availability of datasets in languages other than English, several \cite{Zhang2019,Li2019,Nekvinda2020,Lux2022} researches are focusing on multi-lingual TTS models, which can be trained concurrently in different languages, or that are easily adapted to other languages. Lux and Vu \cite{Lux2022} present techniques to transfer learning from high-resource languages to low-resource languages, using articulatory and phonological features. Zhang et al \cite{Zhang2019} make modifications to the architecture of Tacotron-2 \cite{Shen2018tacotron2}, incorporating a speaker and a language embedding, to develop a model capable of synthesizing audio of different speakers in different languages. Nekvinda and Dusek \cite{Nekvinda2020} also adapt Tacotron-2 to produce a multi-lingual model, which uses different levels of sharing encoder parameters.

Unlike Speech-to-Text (STT) datasets, TTS datasets require good quality audio, preferably recorded in studio with a sample rate of at least 22kHz, and transcripts containing punctuation. Therefore, multilingual datasets for STT systems, such as the Multilingual LibriSpeech (MLS)  \cite{Pratap2020MLS}, Common Voice \cite{Ardila2020commonvoice} or TedX \cite{Salesky2021tedx}, cannot be directly applied in training TTS models. In view of this, the aim of this work is to create a multi-lingual dataset for training TTS models. For that, we present the CML-Multi-Lingual-TTS (CML-TTS) dataset, a version based on MLS dataset, adapted for training TTS models. CML-TTS is derived from readings from LibriVox and consists of audiobooks in seven languages, with a total of 3,233.43 hours and 613 speakers with a sampling rate equal to 24kHz.

This work is organized as follows. Section 2 presents information about CML-TTS, describing its creation process. Section 3 presents details about the YourTTS model. Section 4 shows the results of the experiments performed. The conclusions are presented in the final Section.

\section{CML-TTS}

CML-TTS is a dataset composed of reading audiobooks from the LibriVox\footnote{\url{https://librivox.org/}} project, which uses books from Project Gutenberg\footnote{\url{https://www.gutenberg.org/}}, released in the public domain. In this way, it is possible to make CML-TTS available also in the public domain. It consists of recordings in Dutch, German, French, Italian, Polish, Portuguese, and Spanish, with a sampling rate of 24kHz. The following are details about the CML-TTS creation process. 

%

\subsection{Data Processing Pipeline}\label{section_data_processing_pipeline}

The CML-TTS data processing pipeline consists of four steps. The first step is to download the original audios in mp3, using the LibriVox API, referring to the audiobooks present in the target languages. These languages were selected because they are the same ones present in the MLS dataset, with the exception of English, which was not selected because there is already a large number of datasets available. After the audio files are downloaded, they are converted to wav format with a sample rate of 24kHz, and those with a lower sample rate are discarded.

The second step is to retrive punctuation for each sentences in the MLS, which have no punctuation. For this, the textbooks with punctuation are downloaded. For each sentence $S = \{w_1, w_2, ... , w_n \}$ formed by a sequence of $n$ words $w$, a search is performed for the equivalent sentence $P = \{ wp_1, wp_2, ... , wp_m \}$ formed by a sequence of $m$ words/punctuations $wp$, where $m \geq n$. This is done by defining a search window of length $len$, for each sentence $S$, which is slided through the textbook, word-by-word, in order to find the equivalent sentence $P$. Due to differences in spelling, spaces, hyphenation, etc, it was defined $len$ equal to 90\% of the length of $S$. For sentences comparison, a similarity metric based on Levenshtein's distance is used, normalized between 0 and 1, disregarding punctuation and blank spaces. When finding a segment with similarity $>0.5$ the length of the search window is incrementally increased, word-by-word, until reaching the maximum similarity value. Otherwise, the search window is slided by the rest of the text, and the process is repeated. This entire step is presented in Algorithm \ref{algorithm_search_punc_sentence}.


\begin{algorithm}

\begin{algorithmic} 
\STATE $P \leftarrow$ sentence of the Textbook defined by the search window \;
\IF{$P$ has minimal similarity with $S$}
\STATE Iteratively increase the length of the sentence
\ELSE
\STATE Slide the Textbook search window to define a new sentence $P$
\STATE Repeat the search algorithm for the new sentence $P$
\ENDIF
\caption{Algorithm Search Punctuated Sentence}\label{algorithm_search_punc_sentence}
\end{algorithmic}
\end{algorithm}

Analyzing the LibriTTS dataset, it was verified that the segments have durations between 1 and 20 seconds. However, MLS segments have durations between 10 and 20 seconds, that is, it does not have segments with a duration of fewer than 10 seconds. Therefore, in the third step, segments longer than 15 seconds were divided according to the text punctuation, that is, an audio segment, formed by two text sentences separated by a dot, is divided into two audio segments. This step was performed using Aeneas\footnote{\url{https://www.readbeyond.it/aeneas/docs/index.html}} an audio-text alignment tool.

In this process, the sentence with the greatest similarity may not be the correct sentence spoken in the audio. Or, there may be failures in the audio-text alignment of the segmented sentences, causing an error in the cutting of the audio segments. Therefore, in the last step, validation of the texts is carried out using an STT model to transcribe audio segments and calculate the similarity between the text of the segment and its transcription. The transcription is performed using the Wav2Vec 2.0 XLSR Large \cite{Conneau2021Wav2vec2xlsr}, without any language model, trained originally in 53 languages and fine-tuned individually in each of the languages present in CML-TTS, using the Common Voice dataset \cite{Ardila2020commonvoice} version 6.1.

\begin{figure}[ht!]
    \centering
    \includegraphics[width=8cm]{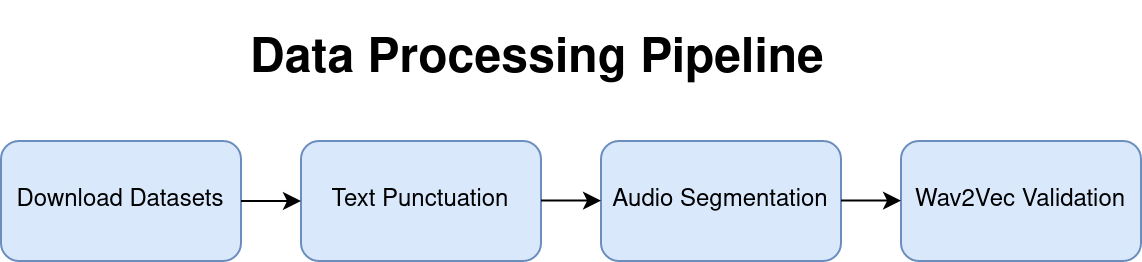}
    \caption{The data processing pipeline is divided into four steps: (1) downloading the original audio; (2) text normalization, adding punctuation; (3) segmentation of the audio in smaller parts; (4) validation of texts through audio transcription.}
    \label{fig_data_processing_pipeline}
\end{figure}

Finally, a similarity metric based on Levenshtein distance is calculated between the transcript and the sentence. The value of this similarity is normalized between 0 and 1, and if the value is less than 0.9, the sentence and the audio are discarded. In this way, a minimum quality of the CML-TTS is guaranteed. The entire Data Processing Pipeline can be seen in Figure \ref{fig_data_processing_pipeline}.

\subsection{CML-TTS Statistics}

Table \ref{ml_libritts_statistics} presents the total duration in hours of each language subsets present in the CML-TTS dataset, and also of the {\it Train}, {\it Test} and {\it Dev} sets. The same {\it Train}, {\it Test}, and {\it Dev} sets of the MLS dataset was kept. In this table, you can also check the duration of the sets in relation to the speaker's gender. A model trained in VoxCeleb 2 \cite{Chung2018voxceleb2} dataset was used for gender classification. 

\begin{table}[ht!]
\centering
\begin{tabular}{c|c|c|c|c|c|c|c|c|c|c|c|c|}
\hline
Language & \multicolumn{6}{c}{Duration} & \multicolumn{6}{c}{Speakers}   \\ \hline
         & \multicolumn{2}{c}{Train} & \multicolumn{2}{c}{Test} & \multicolumn{2}{c}{Dev} & \multicolumn{2}{c}{Train} & \multicolumn{2}{c}{Test} & \multicolumn{2}{c}{Dev}  \\ 
         & M & F & M & F & M & F & M & F & M & F & M & F \\
                    Dutch & 482.82 & 162.17 & 2.46 & 1.29 & 2.24 & 1.67 & 8 & 27 & 3 & 3 & 2 & 4 \\
                   French & 260.08 & 24.04 & 2.48 & 3.55 & 3.31 & 2.72 & 25 & 20 & 8 & 9 & 10 & 8 \\
                   German & 1128.96 & 436.64 & 3.75 & 5.27 & 4.31 & 5.03  & 78 & 90 & 13 & 17 & 13 & 15  \\
                  Italian & 73.78 & 57.51 & 1.47 & 0.85 & 0.40 & 1.52 & 23 & 38 & 5 & 5 & 4 & 6 \\
                   Polish & 30.61 & 8.32 & 0.70 & 0.90 & 0.56 & 0.80 & 4 & 4 & 2 & 2 & 2 & 2 \\
               Portuguese & 23.14 & 44.81 & 0.28 & 0.24 & 0.68 & 0.20 & 20 & 10 & 5 &4 & 6 & 3 \\
                  Spanish & 279.15 & 164.08 & 2.77 & 2.06 & 3.40 & 2.34 & 35 & 42 & 10 & 8 & 11 & 9  \\
  \hline
  Total & \multicolumn{2}{c}{3,176.13} & \multicolumn{2}{c}{28.11} & \multicolumn{2}{c}{29.19} & \multicolumn{2}{c}{424} & \multicolumn{2}{c}{94} & \multicolumn{2}{c}{95} \\ \hline
\end{tabular}
\caption{Total hours and total speakers of Train, Test and Dev sets present in the CML-TTS dataset.}
\label{ml_libritts_statistics}
\end{table}

\begin{figure}[ht!]
    \centering
    \includegraphics[width=10cm]{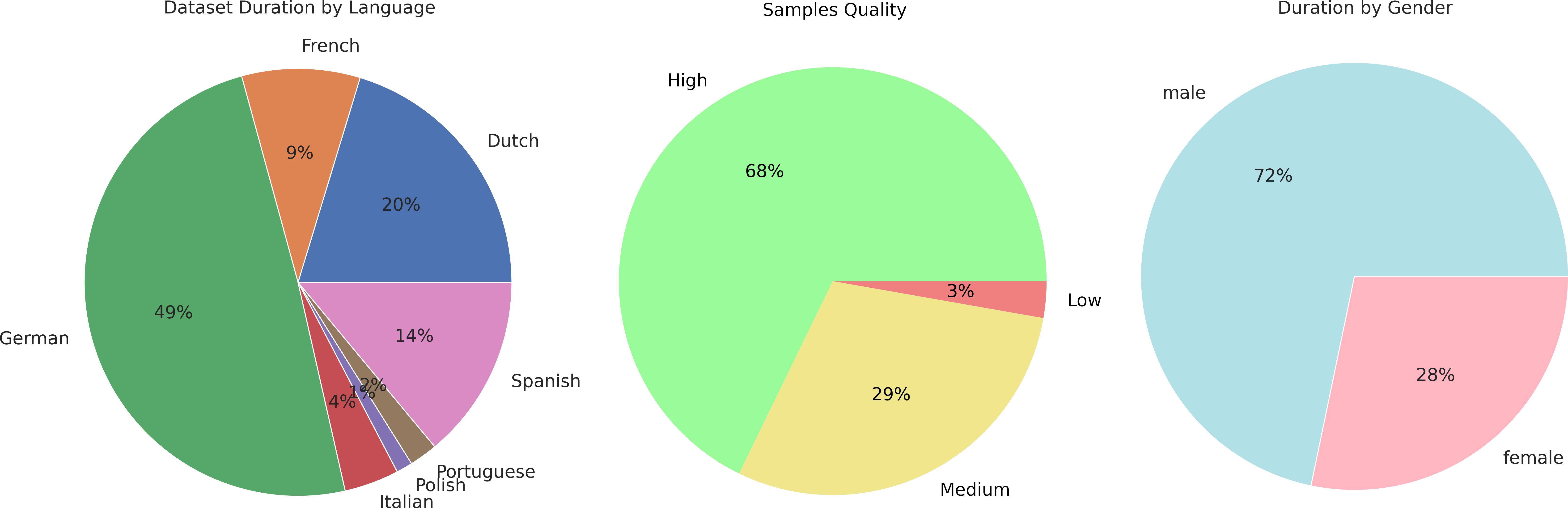}    
    \caption{CML-TTS analysis: on the left, the percentage of the duration of each language in the CML-TTS; in the center, the percentage of samples quality using the WADA; on the right, the percentage of duration in relation to the speaker's gender.}
    \label{fig_data_percentage_languages}
\end{figure}

In Figure \ref{fig_data_percentage_languages} pie charts can be checked indicating the percentage considering the duration of each language (on the left), the percentage of quality of the samples (at center), and the percentage in relation to the gender of the speakers (on the right). The quality of the samples was verified by calculating the SNR using Waveform Amplitude Distribution Analysis (WADA) \cite{Kim2008robust}, and samples with WADA $\geq$ 40dB, 10dB $<$ WADA $<$ 40, and WADA $\leq$ 10dB are indicated respectively as high, medium and low quality.  In the CML-TTS dataset, there are 613 speakers, of which 325 are female and 288 are male. However, when checking the total hours of each gender, the dataset is unbalanced, presenting 2.278 hours for the male gender and 897 hours for the female gender.

%

\begin{figure}[ht!]
    \centering
    \includegraphics[width=9cm]{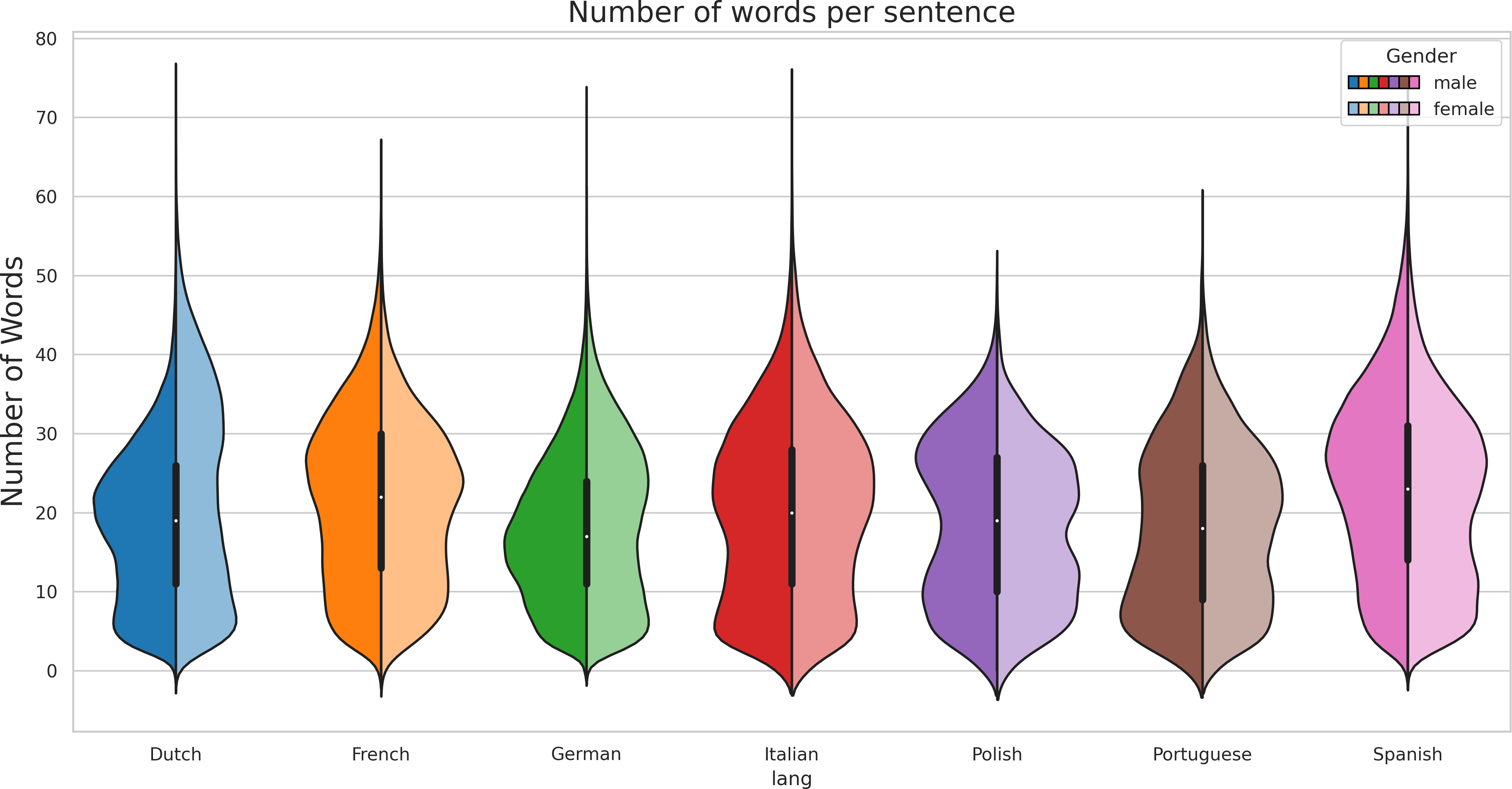}
    \caption{CML-TTS number of words violin plot per language.}
    \label{fig_ml_libritts_num_words}
\end{figure}

Figure \ref{fig_ml_libritts_num_words} shows the violin plot of the number of words per sentence of each of the sub-datasets present in CML-TTS. It can be seen that the sub-datasets are similar in terms of distribution, with an average number of words close to 20. Some differences were caused by the sentence segmentation process and also by the validation process, detailed in Section (\ref{section_data_processing_pipeline}), which discarded some segments. In this figure, it is also noticed that there is a slight difference when analyzing the gender of the speakers, but it does not affect the quality of the dataset.

\begin{figure}[ht!]
    \centering
    \includegraphics[width=9cm]{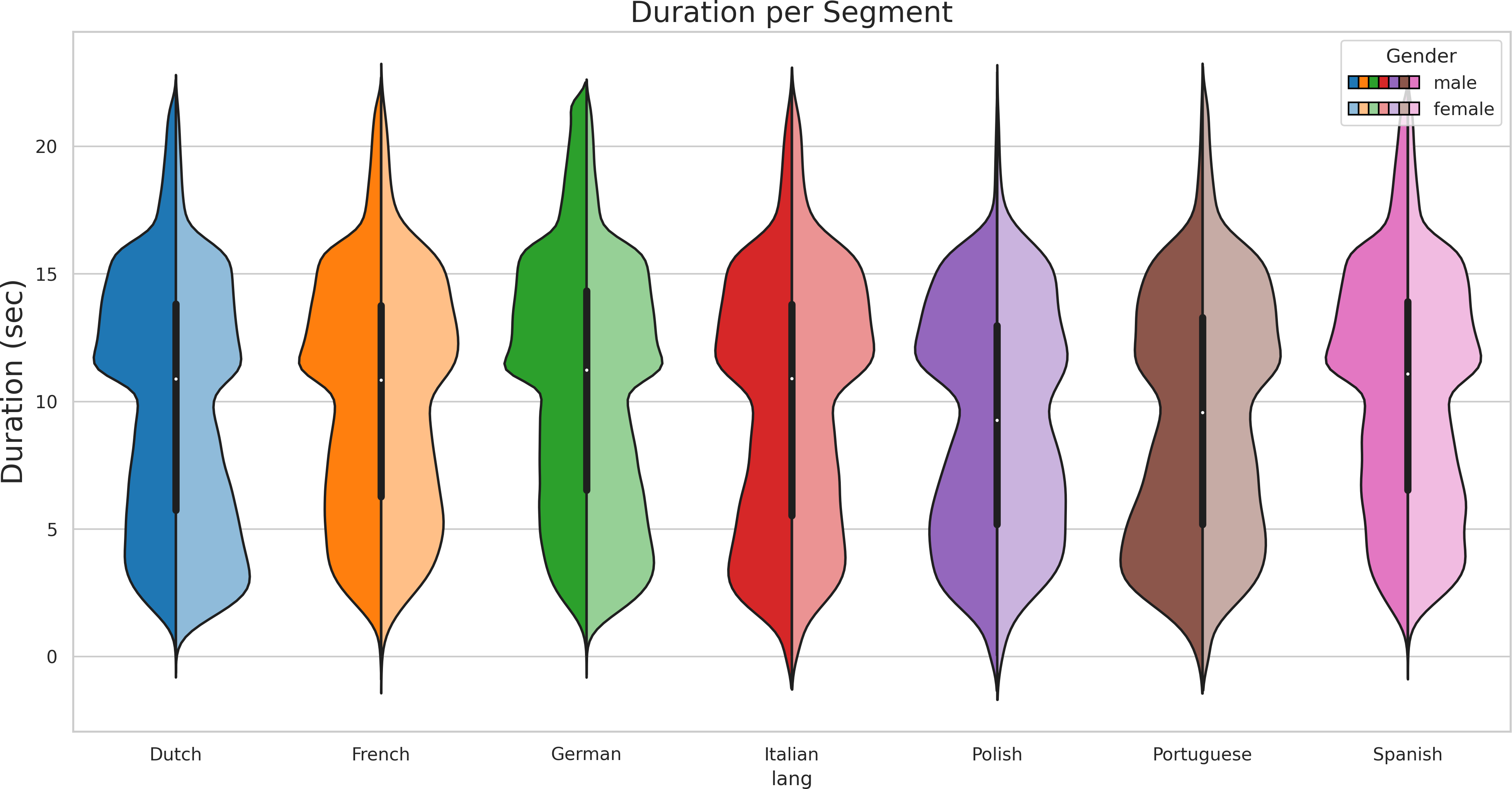}
    \caption{CML-TTS segments duration violin plot per language.}
    \label{fig_ml_libritts_duration}
\end{figure}

Figure \ref{fig_ml_libritts_duration} shows the violin plot of the duration per sentence of each of the sub-datasets present in CML-TTS. It can be seen that the segmentation step was effective, making the segments last between 1 and 22 seconds. Each of the sub-datasets has an average sentence length of between 8 and 12 seconds. However, a disadvantage of the segmentation and validation process is that it made the dataset unbalanced in terms of the duration of the segments, with a greater number of segments with durations close to 12s. In this figure, it is also verified that there is a slight difference in relation to the gender of the speakers.

\begin{figure}[ht!]
    \centering
    \includegraphics[width=6cm]{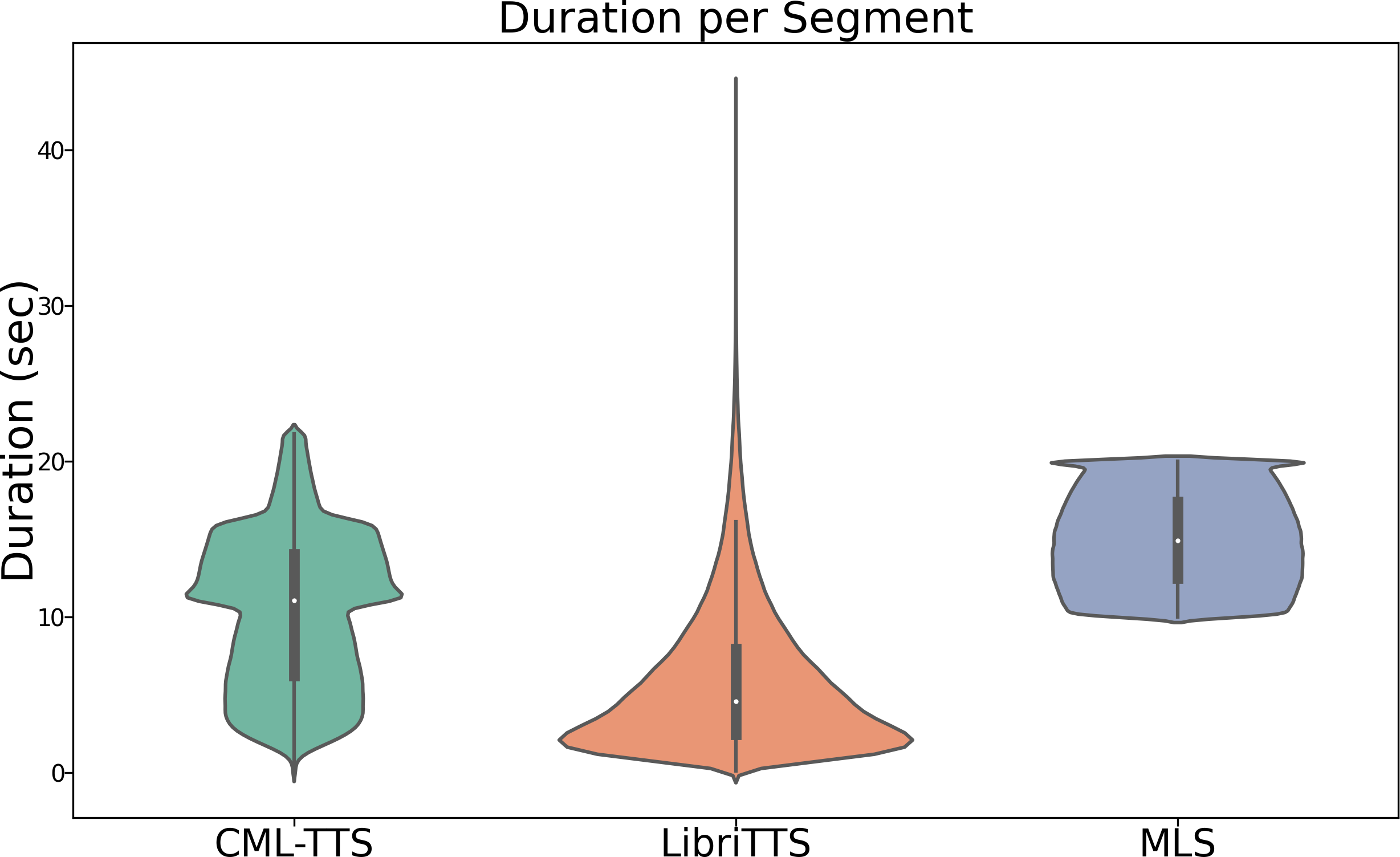}
    \includegraphics[width=6cm]{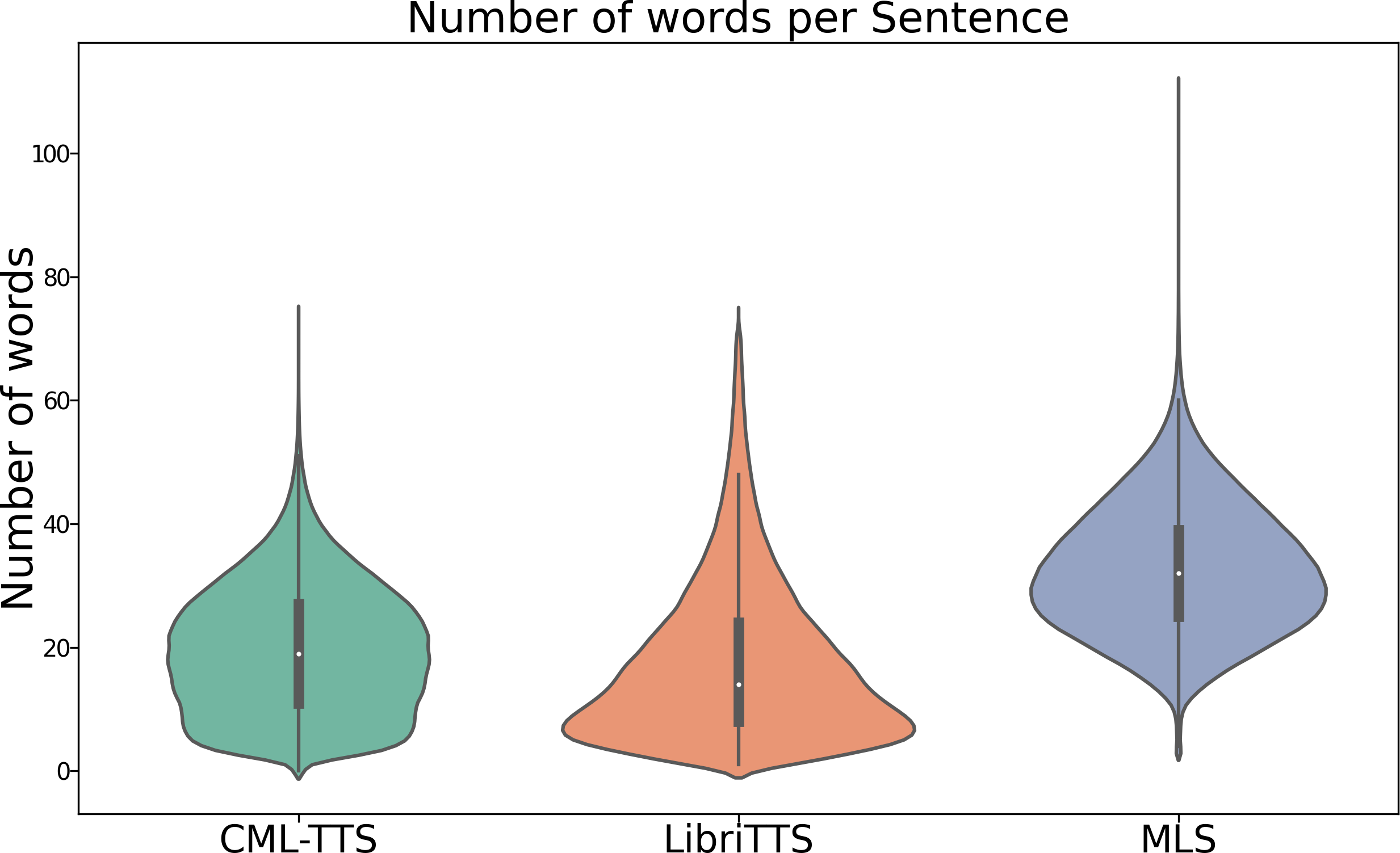}
    \caption{Comparison among CML-TTS, LibriTTS, and MLS. On the left: durations of segments; right: number of words per sentence.}
    \label{fig_comparative_datasets}
\end{figure}

Duration diversity is required for training successfully most TTS models. For this reason, 
it is interesting to analyze how our dataset is near or far from LibriTTS. Therefore, Figure \ref{fig_comparative_datasets}, on the left, shows a comparison of the duration of the segments between the CML-TTS, LibriTTS, and MLS datasets. It can be seen that most of the LibriTTS segments have a duration of less than 10s, while the MLS segments are distributed between 10 and 20s. The CML-TTS segments are distributed between 0 and 20s. There was a decrease in the average duration of the segments, which was originally 18s in MLS, decreasing to 12s in CML-TTS, while the average in LibriTTS is closer to 2s. Due to the sentence validation process, there was a drastic reduction in the total number of segments with duration close to 10 seconds. 

Also looking at LibriTTS, very long sentences with a large number of words are not desired. Figure \ref{fig_comparative_datasets}, on the right, shows a comparison of the number of words in sentences between the CML-TTS, LibriTTS, and MLS datasets. With the segmentation process, similar to the length of the audio segments, there was also a reduction in the average number of words in the sentences, falling from 35 in the MLS to 20 in the CML-TTS. Therefore, the segmentation process made CML-TTS closer to LibriTTS, which averages 10 words per sentence.


\section{YourTTS Model}

YourTTS \cite{Casanova2022yourtts} is a multilingual zero-shot multi-speaker TTS model, that was built upon VITS \cite{Kim2021vits} architecture. The goal of zero-shot multi-speaker TTS models is to generate speech of speakers not seen during training, employing only a few seconds of the target speaker's voice. Although the YourTTS model was proposed on the zero-shot multi-speaker scenario and the main objective of the multilingual training was to reduce the number of speakers needed to develop a zero-shot multi-speaker TTS model in a target language. YourTTS multilingual results are very good and it can do cross-lingual speaker transfer with good quality as well. In addition, the model can generate high-quality speech in the 3 languages that the model was trained in. Although the authors did not explore code-switching \cite{Nekvinda2020}, YourTTS can do code-switching with great quality producing the word of the other language with naturality and great speaker similarity even for speakers not seen in training.

YourTTS architecture is composed of a Text-Encoder, Posterior-Encoder, Duration Predictor, Flow-Based Decoder, Vocoder, and a pre-trained Speaker-Encoder.

Text-Encoder has Transformer-based architecture, as used in \cite{Kim2020GlowTTS,Kim2021vits,Casanova2021SCGlowTTS} consisting of 10 transformer blocks and 196 hidden channels. For multilingual training, the authors concatenated 4-dimensional trainable language embeddings into the embeddings of each input character.

The Posterior-Encoder, proposed by \cite{Kim2021vits}, is responsible for connecting the Flow-Based to the Vocoder during the training or voice conversion mode. Its architecture is composed of 16 non-causal residual blocks, as per \cite{Oord2016WaveNet}. The Posterior-Encode receives as input a linear spectrogram and speaker embedding and produces latent variable $z$, which is an intermediate representation, similar to a mel-spectrogram, however, here this representation is learned by the model. During inference, the  Posterior-Encoder is not used and the latent variable $z$ is predicted by the Flow-Based Decoder.

As Vocoder it uses the HiFi-GAN model \cite{Kong2020hifigan} with the discriminator modifications introduced by \cite{Kim2021vits}.

The pre-trained Speaker-Encoder is based on the H/ASP \cite{Heo2020Clova} architecture and was trained with the Prototypical Angular \cite{Chung2020} plus Softmax loss functions in the VoxCeleb 2 \cite{Chung2018voxceleb2} dataset. The Speaker-Encoder was used to extract speaker embeddings that were conditioned on the Duration Predictor, Flow-Based Decoder, and Vocoder.

\section{Experiments}

This section describes the experiments performed using the YourTTS model with CML-TTS and LibriTTS. The LibriTTS dataset was included in order to improve the generalization of the model for speakers not seen during training because LibriTTS has a large number of speakers, which is approximately 1100 speakers. Thus, the model was trained in eight languages: Dutch, English, French, German, Italian, Polish, Portuguese, and Spanish.

\subsection{Training}

YourTTS was trained using the Train subset of the CML-TTS and LibriTTS datasets (clean-100 and clean-360). To speed up the training, we carry out transfer learning from the official checkpoints, trained in three languages: English, Portuguese and French, using the VCTK \cite{Yamagishi2016Vctk}, TTS-Portuguese \cite{Casanova2022tts} and M-AILABS datasets \cite{Mailabs2017}. The training was performed on a DGX-A100, with 82G of memory, for a maximum period of two weeks, using the AdamW optimizer, with a learning rate equal to 0.001, betas equal to 0.8 and 0.99, and batch size equal to 60.

As Speaker-Encoder, we followed the YourTTS paper and uses the H/ASP \cite{Heo2020Clova} pre-trained using the VoxCeleb2 \cite{Chung2018voxceleb2} dataset with Prototypical Angular \cite{Chung2020} together with Softmax as loss function. The embeddings were previously extracted to save computational resources.


\subsection{Results}

The objective of these experiments is to evaluate whether the CML-TTS has sufficient quality for training TTS models, therefore, we only evaluate the languages present in the CML-TTS dataset, since the use of LibriTTS was only to increase the generalization power of the model due to a large number of different speakers. 

To evaluate the similarity between the synthesized speech and the ground truth, we calculate the Speaker Encoder Cosine Similarity (SECS)  \cite{Casanova2022yourtts} between the embeddings extracted from the generated audios and from the ground truth audios. In SECS, the closer to 1, the greater the similarity, while the closer to -1 indicates low speaker similarity.

The embeddings were extracted using the model proposed by \cite{Wan2018ge2e} trained on the VoxCeleb 1, 2 datasets \cite{Nagrani2017VoxCeleb,Chung2018voxceleb2}. The implementation is available in the Resemblyzer package \cite{Jemine2019}, which is the same one used in the YourTTS paper, in order to allow a fair comparison. 

A total of 1,000 sentences for each language were synthesized using different speakers. Since some languages in the CML-TTS have a Test set with fewer than a thousand samples, the sentences were randomly extracted from the Dev and Test sets, which were not used during the training phase. Ten speakers were selected from the Train set and another ten from the Test and Dev sets, or all speakers when the total was less than ten. The evaluation process using unseen speakers is named the zero-shot way. 

Table \ref{tab_secs_results} shows the results of the SECS metric of seen and unseen speakers during training for each of the languages present in CML-TTS. The worst result was obtained in the Portuguese language, however, in the original paper YourTTS presented SECS values for the Portuguese language equal to 0.740, in the worst experiment, and 0.798 in the best experiment. Our experiment for the Portuguese language presented a result of the SECS metric equal to 0.777.

\begin{table}[ht!]
  \begin{center}
    \begin{tabular}{l|c|c|} 
      \multirow{2}{*}{Language} & \multicolumn{2}{c}{SECS}   \\
                                & Seen & Unseen  \\
      \hline
      Dutch & 0.8181 $\pm$ 0.0050 & 0.8010 $\pm$ 0.0042 \\
      French & 0.8299 $\pm$ 0.0044& 0.8205 $\pm$ 0.0044 \\
      German & 0.8366 $\pm$ 0.0046 & 0.8453 $\pm$ 0.0032 \\
      Italian & 0.7904 $\pm$ 0.0069 & 0.7906 $\pm$ 0.0059 \\
      Polish & 0.8280 $\pm$ 0.0039 & 0.8041 $\pm$ 0.0043 \\
      Portuguese & 0.7772 $\pm$ 0.0063 & 0.7717 $\pm$ 0.0079 \\
      Spanish & 0.8536 $\pm$ 0.0048 & 0.8180 $\pm$ 0.0050 \\
    \end{tabular}
  \end{center}
  \caption{Speaker Encoder Cosine Similarity (SECS) }
  \label{tab_secs_results}
\end{table}

To verify if the synthesized sentences were in agreement with the input text, we performed the transcription of the synthesized audios using a Speech-to-Text model. The model chosen was Wav2Vec 2.0 XLSR Large \cite{Conneau2021Wav2vec2xlsr}, the same used for validation during the data processing pipeline in the Section (\ref{section_data_processing_pipeline}). Using the transcripts and the ground truth text, we calculated the Word Error Rate (WER) and Character Error Rate (CER) metrics. CER metric is calculated according to the equation $CER = \frac{S + D + I}{N}$ where $S$ is the number of substitutions, $D$ deletions, $I$ insertions, and $N$ is the total characters of the ground truth text. WER metric operates is similar, but operating at the word level instead. 

Table \ref{tab_wer_cer_results} shows the results of the WER and CER metrics of seen and unseen speakers during training for each of the languages present in CML-TTS. In both metrics, the language that presented the best results was Spanish, while Portuguese had the worst results. However, this is explained because in Portuguese there was a significant change in orthographic rules due to an Orthographic Agreement\footnote{\url{https://www.instituto-camoes.pt/en/activity-camoes/what-we-do/teach-portuguese/orthographic-agreement}} carried out in 1990, to unify orthography among Portuguese-speaking countries. Therefore, this influenced the results, given that the books present in LibriVox were published prior to this agreement. It should also be considered that a language model was not used to transcribe the audios.

\begin{table}[ht!]
  \begin{center}
    \begin{tabular}{l|c|c|c|c|} 
      \multirow{2}{*}{Language} & \multicolumn{2}{c}{WER}  & \multicolumn{2}{c}{CER}  \\
                                & Seen & Unseen & Seen & Unseen  \\
      \hline
      Dutch &  0.3223 & 0.3062 & 0.1192 & 0.0941\\
      French &  0.2909 & 0.1636 & 0.1330 & 0.0490\\
      German &  0.2305 & 0.1737 & 0.0830 & 0.0407\\
      Italian & 0.2956 & 0.2002 & 0.1260 & 0.0574 \\
      Polish & 0.3133 & 0.3285 & 0.0955 & 0.0970 \\
      Portuguese & 0.4134 & 0.4548 & 0.1985 & 0.2098 \\
      Spanish &  0.1862 & 0.1039 & 0.07947 & 0.0243\\
    \end{tabular}
  \end{center}
\caption{WER and CER metrics between ground truth sentences and transcriptions using Wav2Vec.}
\label{tab_wer_cer_results}
\end{table}

\section{Conclusions}\label{sec:conc}
We have presented the CML-Multi-Lingual-TTS dataset, a dataset composed of Librivox recordings, comprising audiobooks in seven languages: Dutch, German, French, Italian, Polish, Portuguese and Spanish. This dataset was created from the MLS dataset, performing a data processing step in order to make it more suitable for training TTS models. Statistical data of the dataset were presented, comparing it with the main dataset for training TTS models, LibriTTS. Experiments were also performed using the YourTTS model, training it with the CML-TTS and LibriTTS datasets, demonstrating that the dataset is suitable for training TTS models.

\section{Acknowledgements}


The authors are grateful to CEIA at UFG for their support and to Coqui and CyberLabs for their valuable assistance. We also thank the LibriVox volunteers for making this project possible.

%

\bibliographystyle{splncs04}
\bibliography{paper}

\end{document}